\documentclass[conference]{IEEEtran}
\IEEEoverridecommandlockouts
\usepackage{cite}
\usepackage{amsmath,amssymb,amsfonts}
\usepackage{algorithmic}
\usepackage{graphicx}
\usepackage{textcomp}
\usepackage{subfigure}

\def\BibTeX{{\rm B\kern-.05em{\sc i\kern-.025em b}\kern-.08em
    T\kern-.1667em\lower.7ex\hbox{E}\kern-.125emX}}
 
\begin{document}

\title{A Hierarchical Terminal Recognition Approach based on Network Traffic Analysis\\
\thanks{}
}
\author{\IEEEauthorblockN{Lingzi Kong, Daoqi Han, Junmei Ding, Mingrui Fan and Yueming Lu*}
\IEEEauthorblockA{School of Cyberspace Security, Beijing University of Posts and Telecommunications, Beijing, China}
\IEEEauthorblockA{\{klz, handq, jmding, fanmingrui, ymlu\}@bupt.edu.cn}
\IEEEauthorblockA{*Corresponding Author}
}

\maketitle

\begin{abstract}
Recognizing the type of connected devices to a network helps to perform security policies. In smart grids, identifying massive number of grid metering terminals based on network traffic analysis is almost blank and existing research has not proposed a targeted end-to-end model to solve the flow classification problem. Therefore, we proposed a hierarchical terminal recognition approach that applies the details of grid data. We have formed a two-level model structure by segmenting the grid data, which uses the statistical characteristics of network traffic and the specific behavior characteristics of grid metering terminals. Moreover, through the selection and reconstruction of features, we combine three algorithms to achieve accurate identification of terminal types that transmit network traffic. We conduct extensive experiments on a real dataset containing three types of grid metering terminals, and the results show that our research has improved performance compared to common recognition models. The combination of an autoencoder, K-Means and GradientBoost algorithm achieved the best recognition rate with F1 value of 98.3$\%$.

\end{abstract}


\section{Introduction}

Network traffic analysis has been a topic of interest from the
early stage of the Internet, and has received the most attentions in Internet of Things (IoT) in order to manage the overall performance of their network. With the thriving of intelligent applications, the antiquated power infrastructure that delivers power to consumers is progressively replaced by a series of digital systems, which is referred to as the smart grid (SG). The smart grid is a modernized digital power grid consisting of smart grid systems, Advanced Metering Infrastructure (AMI) \cite{b1}, smart protection control, etc. Among them, the grid metering terminal is an important part of AMI which can greatly assist utility providers (UPs) to monitor \cite{b2}, control and forecast the energy consumption, increasing the quality of the power.

Grid metering terminals are intelligent terminals in the perception layer of the IoT. The development of IoTs is driving technology and computing ability into the smallest devices, making data collection and availability ubiquitous \cite{b3}. As the basic equipment for smart grid data collection, grid metering terminals are mainly used to collect, measure and transmit data. Furthermore, in order to adapt to the use of smart grids and new energy sources, these terminals also have many functions such as user-side control function, anti-theft and other intelligent functions. The information for those functions is passed through network traffic. Therefore, making good use of this information to accurately identify the type of grid metering terminals is important for maintaining smart grid. Meanwhile, recognizing the type of the connected device to a network helps to perform security policies as well as ensuring accurate functionality of the device \cite{b4}, and it is also a significant challenge for improving grid operation efficiency.

As for smart grid data research, most studies focus on typical attacks \cite{b5,b6} and failures\cite{b7,b8}, ignoring the management of large daily network traffic. Due to the particularity of smart grid data, the analysis methods are divided into physical state-based, game theory-based and data-driven. The method based on the physical state \cite{b9} utilizes the physical nature of the power grid. Although the accuracy rate is high, it requires additional investment and is low cost-effective. The game theory-based method \cite{b10} conducts simulated games for possible anomalies, whereas this method is only theoretically derived and has not been practically verified. Data-driven analysis methods maximize the advantages of smart grids. This approach models behavior or type by analyzing the data flowing through terminals. 
\begin{figure}[b]
\centerline{\includegraphics[scale=0.28]{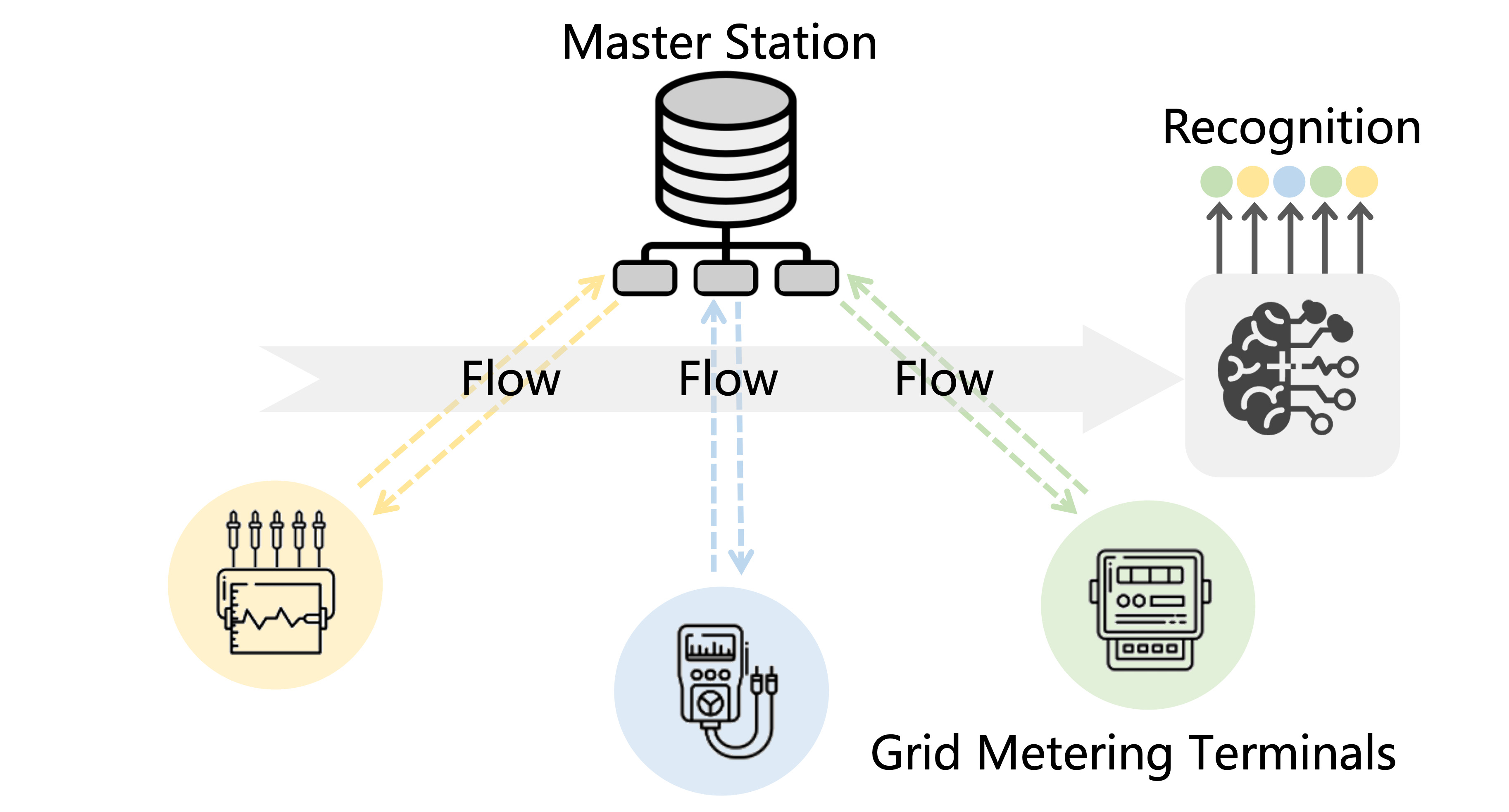}}
\caption{The grid system architecture.}
\label{fig1}
\end{figure}

Therefore, we consider applying network traffic analysis technology to the field of smart grids to recognize the type of grid metering terminals. 
The smart grid data is different from ordinary network traffic. As shown in Fig.\ref{fig1}, many types of grid metering terminals interact with one or several master stations in an area (usually only one master station), forming a central network. The master station receives massive data flows. Recognizing the type of grid metering terminals by analyzing network flows is of great significance for smart metering infrastructure management. L. Lu et al. \cite{b11} designed a cognitive memory-guided autoencoder model which excavates the latent space feature representation through autoencoder. V. Labayen et al. \cite{b12} studied real-time traffic classification. Their model obtained the hidden information of the target time window through clustering, and achieved good performance.

Inspired by \cite{b11,b12}, we propose a grid metering terminal recognition approach. The particular aspects of the proposed approach are driven by the specifics of the smart grid systems. The approach firstly split each flow into segments, which is important for analyzing flows at a deeper level. Secondly, we design a set of modules including an autoencoder, a clustering algorithm and a classification algorithm. We use an autoencoder to extract better representations of behavioral features, a clustering method to cluster and encode segment features of each flow to generate potential feature representations and  a classification algorithm to recognize terminals type. In summary, the listed content is the main contribution of this work:

\begin{itemize}
\item We design a hierarchical structure combined with flow features and fine-grained segmented features to analyze network traffic.
\item We propose an end-to-end model by mining compressed representation of terminal behavior, distinguishing packets sending rhythms of different segments to recognize the type of grid metering terminals successfully.
\item We conduct extensive experiments on a real grid metering terminal dataset. The experimental analysis demonstrates that the encoded features produced by the hierarchical model contribute to the improvement of recognition performance.
\end{itemize}

\section{Related Work}

\subsection{Smart Grid Data Analysis}

Analyses of smart grid in academia mainly focus on typical attack behaviors \cite{b13,b14,b28}, including intrusion detection harming the confidentiality, integrity, and availability \cite{b5,b6,b15}. Shrestha et al. \cite{b3} proposed a methodology for security classification applied to smart grid infrastructures, which looks more at the functionalities of the system than at the possible attacks. In addition, the research on smart grid information also involves fault detection and analysis of the smart grid. Calderaro et al. \cite{b7} detect failures in data transmission and faults in the distribution network with the help of Petri nets analysis and matrix operations. Reference \cite{b8} used a network inference algorithm based on Markov random fields and dependency graphs to study fault detection and localization in transmission lines. However, all those approaches are based on the powerline level (i.e. modeling and observing individual relays \cite{b7} or phasor angles across the transmission line buses \cite{b8}), not the traffic flows above it. Another part of the study analyzed user behavior based on data from grid metering terminals, Zeifman et al. \cite{b17} clustered user behaviors based on household electricity consumption data, Yip et al. \cite{b2} identified electricity stealing behaviors, and Rossi \cite{b19} clustered more extensive abnormal behaviors of grid metering terminals.

By summarizing the above work, we find that there is no research on the recognition of terminal types for smart grids. Therefore, we consider recognizing the types of grid metering terminals in the smart grid infrastructure. We analyze the difference between various terminals from a data perspective rather than a system structure perspective by learning from flows generated by them in smart grids. In addition, we focus on the traffic characteristics of normal terminals rather than attack behaviors to achieve better management of smart grid systems.

\subsection{Network Traffic Classification}

Both academic researchers and industries have proposed and developed many Network Traffic Classification (NTC) strategies in the last decades. The most common methods for NTC are based on Port numbers and Deep Packet Inspection(DPI), as well as on Statistics-driven classification. The port number-based classification method compares the port number information in the header of the transport layer packet with the known specific application layer port number to classify network traffic. Since the extraction process is simple and fast, it is commonly used in firewalls and access control lists (ACLs) \cite{b20} . However, the widespread use of techniques such as port obfuscation, NAT (network address translation), and random port assignment make this method unreliable. DPI relies on payload inspection to find out well-known patterns that are able to univocally characterize the different traffic types. The shortcomings of this method are mainly that accessing the full payload is computationally expensive, and the encrypted traffic cannot be identified due to privacy protection issues \cite{b23}.

Statistic-driven approaches rely on high-level information extracted from the packets header. Moore \cite{b24} proposed 248 traffic features; Labayen et al. \cite{b12} used random forest and K-means methods to classify five behavioral traffic flows. Crotti et al. \cite{b25} proposed a protocol fingerprint based on the probability density of inter-packet arrival times to classify HTTP, Mail Protocol 3 (POP3) and Simple Mail Transfer Protocol (SMTP) with 91$\%$ accuracy; Wang \cite{b26} applied deep learning to network traffic classification for the first time, building a stacked autoencoder (SAE) classification model and using the first 1000 bytes of the TCP payload as the model input to classify 58 protocols such as HTTP and SMTP. Lotfollahi et al. \cite{b23} proposed a "Deep Packet" method to classify applications and network traffic, which takes the first 1500 bytes of IP packets as input data and builds SAE and one-dimensional CNN models. D’Angelo et al. \cite{b27} proposed six combined models based on convolutional and recurrent neural networks, autoencoder and other models to mine the spatiotemporal features of network traffic, proving the effectiveness of combining deep learning with statistical features. 

Based on the above work, we believe that it is worthwhile to use the application layer load and statistical features as the combined features, and to adopt the clustering method, classification method of machine learning and deep learning at the same time. Therefore, we propose an approach for the identification of grid metering terminals, which uses a combination of behavioral features and statistical features to construct a hierarchical model. Our model includes an autoencoder, a clustering method and a classification method to identify the types of network flows, achieving better results than common models.

\section{Problem Formulation}
\subsection{Flow}
Network traffic is the data flowing in the network, containing many data packets. As it is widely known, network traffic observable at a specific tapping point results from the composition of many individual flows.  Each flow is defined as the traffic characterized by having a unique bi-directional combination of a specific source IP address, destination IP address, source Port, destination Port, and transport layer protocol (TCP, UDP, etc.), and this combined characteristic is named as five-tuple. The network traffic we studied records the information transmission between grid metering terminals and the master station. 

From Fig.\ref{fig1} we can see that the data we analyze is a central network rather than a traditional social network. We denote a flow as \(F_n\), and if \(F_n\) is sent and received by a terminal of type \(C\), the type of flow \(F_n\) is \(C\), where \(C\in \mathbf{C} \) and \(\mathbf{C}\) represents all terminal types in the system. \(F_n\) is composed of \(E\) data packets \(\left \{ p_1,p_2,\cdots ,p_E\right \}\), where \(p_e\) represents the \(e\)th packet, containing basic characteristics such as data length, source IP, timestamp, etc. The basic characteristics of each packet can be extracted through the network packet parsing method. Based on the above basic characteristics, we generated a set of features for each flow. The flow features of \(F_n\) can be denoted as \(x_{n}^{FF}\in \mathbb{R^{D}}\), where \(D\) is the feature dimension. 
\subsection{Segment}
A segment is a portion of a flow intercepted by time. More precisely, the sequence of packets in a flow is split into shorter temporal windows, allowing the feature of the flow exhibited in this set of segments. We suppose flow \(F_n\) contains \(L\) segments of equal length \(\tau\), where \(\tau\) is a hyper-parameter which indicates a certain time window size (e.g. 5 minutes). We then have \(F_n=\left \{S_1,S_2,\cdots ,S_L\right \}\), where \(S_l\) is the \(l\)th segment containing \(M_l\) continuous packets.

In this structure, features are extracted by segment. The segment-wise features of \(F_n\) can also be denoted as \(X_{n}^{SF}=\left \{s_{n1}^{SF},s_{n2}^{SF},\cdots,s_{nL}^{SF} \right \}\in \mathbb{R}^{L\times G}\) where \(L\) is the number of segments, \(s_{nl}^{SF}\) is a \(G\)-dimensional vector which indicates the features of segment \(S_l\). In addition, the segment-wise flow features are high-dimensional features that will be reconstructed by our model, and we use embeddings to represent vectors generated after feature encoding.
\subsection{Problem}
Let \(\left \{F_{1},F_{2},\cdots ,F_{N}\right \}\) denote some network flows, and a set of terminal types \(\mathbf{y}=\left \{y_{1},y_{2},\cdots ,y_{N}\right \}\) where \(y_n\) is the ground-truth type of flow \(F_n\). Our task is to recognize the terminal type of the flow based on it's features, getting the predicted type \(\hat{\mathbf{y}}=\left \{\hat{y}_{1},\hat{y}_{2},\cdots ,\hat{y}_{N}\right \}\). The goal of our model is to minimize the classification error between \(\mathbf{y}\) and \(\hat{\mathbf{y}}\). The basic information of network flows is extracted from the original packet capture (PCAP) traffic data. The features of each network flow are extracted from both the flow itself and segments. In this work, we aim to recognize the type of the terminal to which the flow \(F_n\) belongs via mining hidden representations of the network flow and using multiple machine learning algorithms effectively. As for the methods, we consider cluster algorithm, classification algorithm and autoencoder technology.

\section{Proposed Model}
In this section, we propose a new network traffic recognition framework for grid metering data to recognize the type of each flow’s terminal. We first describe the extracted features from the two dimensions of flow and segment (section \ref{A}), then we construct the hierarchical data structure (section \ref{B}), and finally we elaborate on our proposed model (section \ref{C}).
\subsection{Feature Selection}\label{A}
Unlike traditional network traffic, the problem we aim to solve is to classify the network traffic generated by grid metering terminals. Therefore, we excavated a set of features from the two aspects of statistical characteristics and communication specifications on behalf of terminals' behavior. 
\begin{table}[b]
\caption{Feature Set}
\begin{center}
\resizebox{\linewidth}{!}{
\begin{tabular}{ccc}
\hline
\textbf{Feature Category}&\textbf{Feature}&\textbf{Dimension} \\
\hline
Statistical Feature&Packet Count&2\\
&Packet Count Send-Receive Ratio&1\\
&Sum of Packet Size&2\\
&Average Packet Size&2\\
&Standard Deviation of Packet Size&2\\
&Packet Size Sum Sent-Receive Ratio&1\\
&Range of Packet Size &2\\
&Average of Packet Inter-arrival Time&2\\
&Standard Deviation of Packet Inter-arrival Time &2\\
\hline
Behavioral Feature&Test&4\\
&Write&2\\
&Identification&2\\
&Read Operations&12\\
&Transport&6\\
&Individuation&2\\
\hline
\end{tabular}}
\label{tab1}
\end{center}
\end{table}

\begin{figure*}[t]
\centering
\subfigure{a)\label{2 a)}
\includegraphics[width=5.1cm,height=3.4cm]{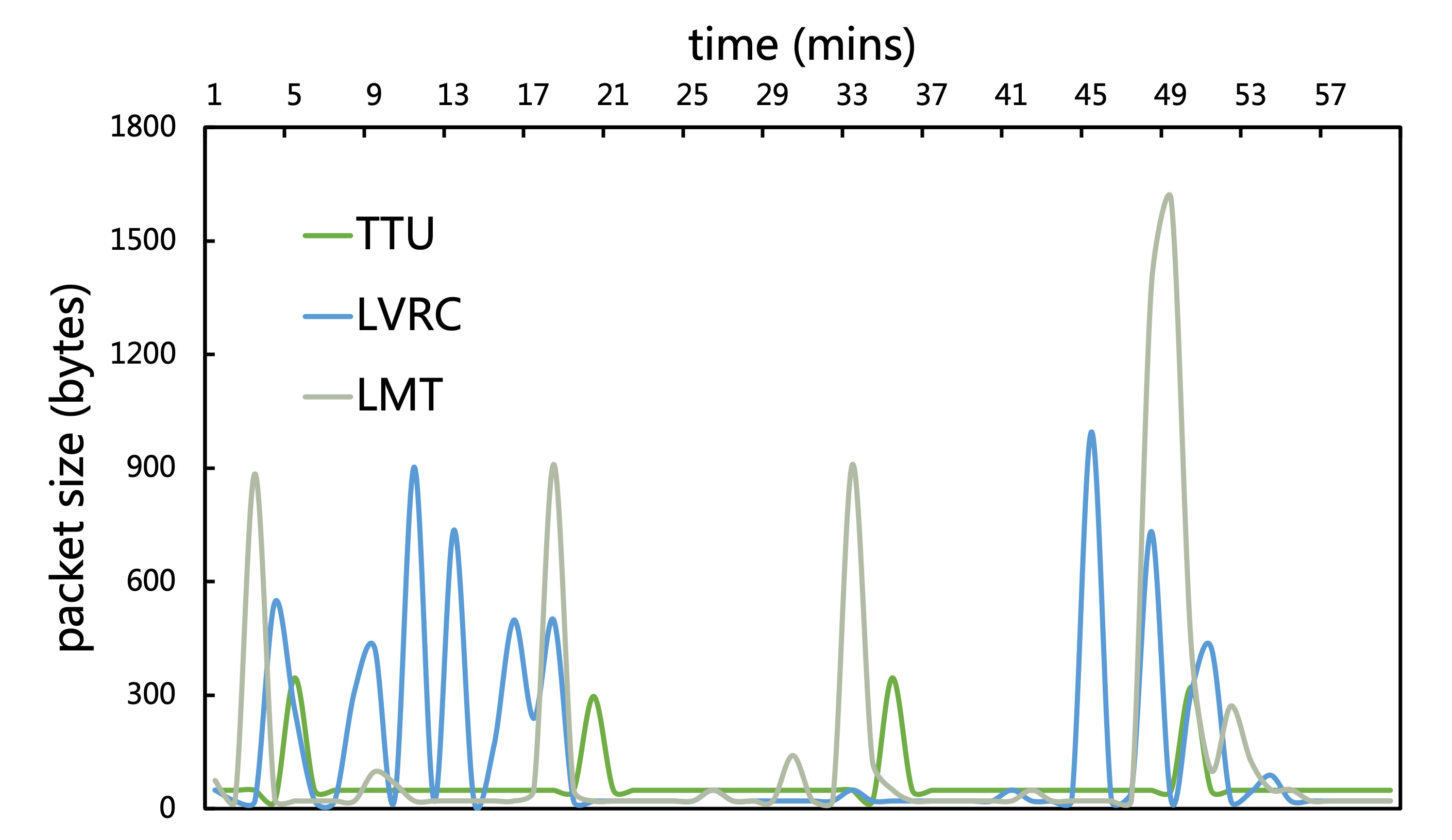}
}\subfigure{b)\label{2 b)}
\includegraphics[width=3.6cm,height=3.4cm]{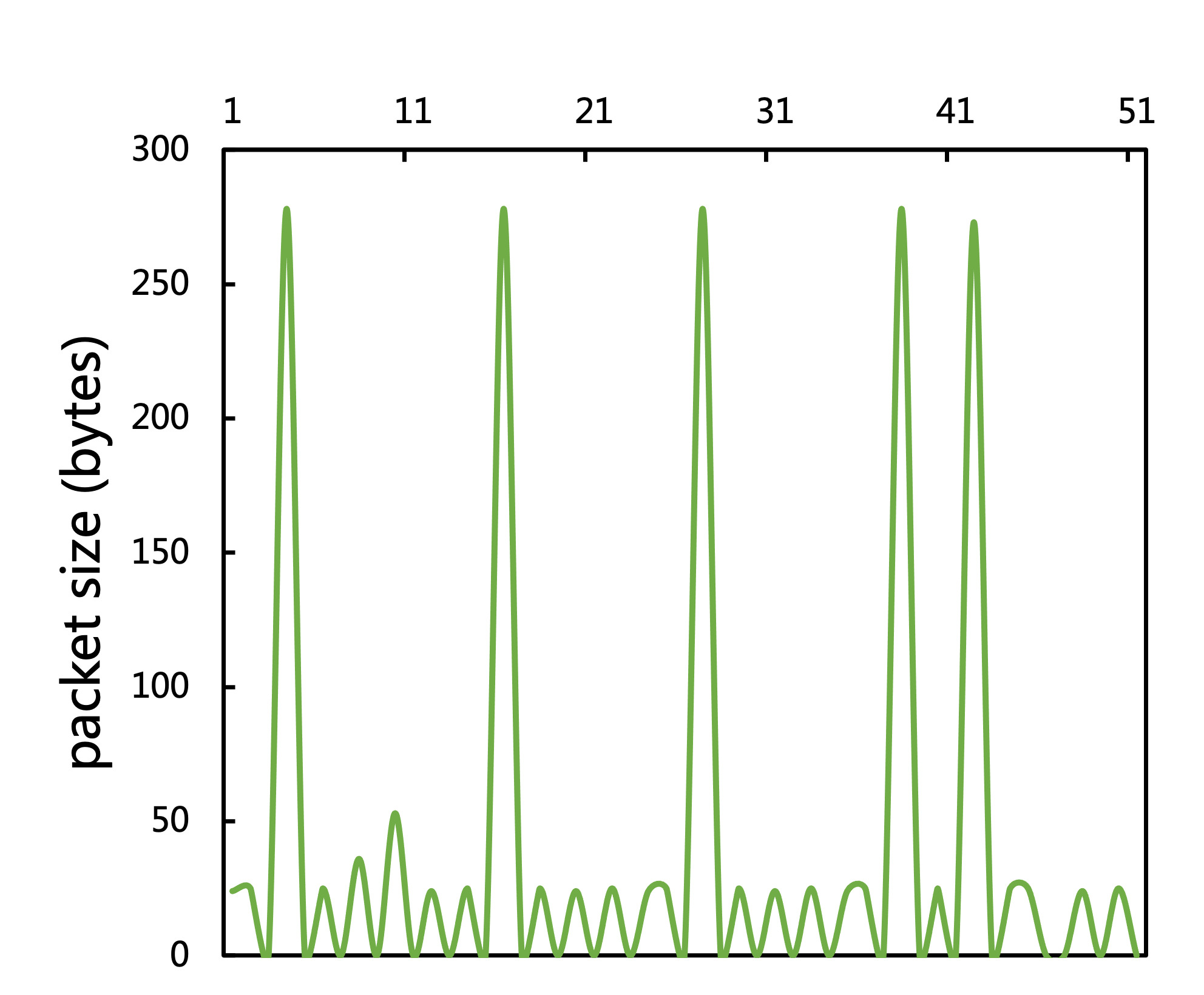}
}\subfigure{c)\label{2 c)}
\includegraphics[width=3.6cm,height=3.4cm]{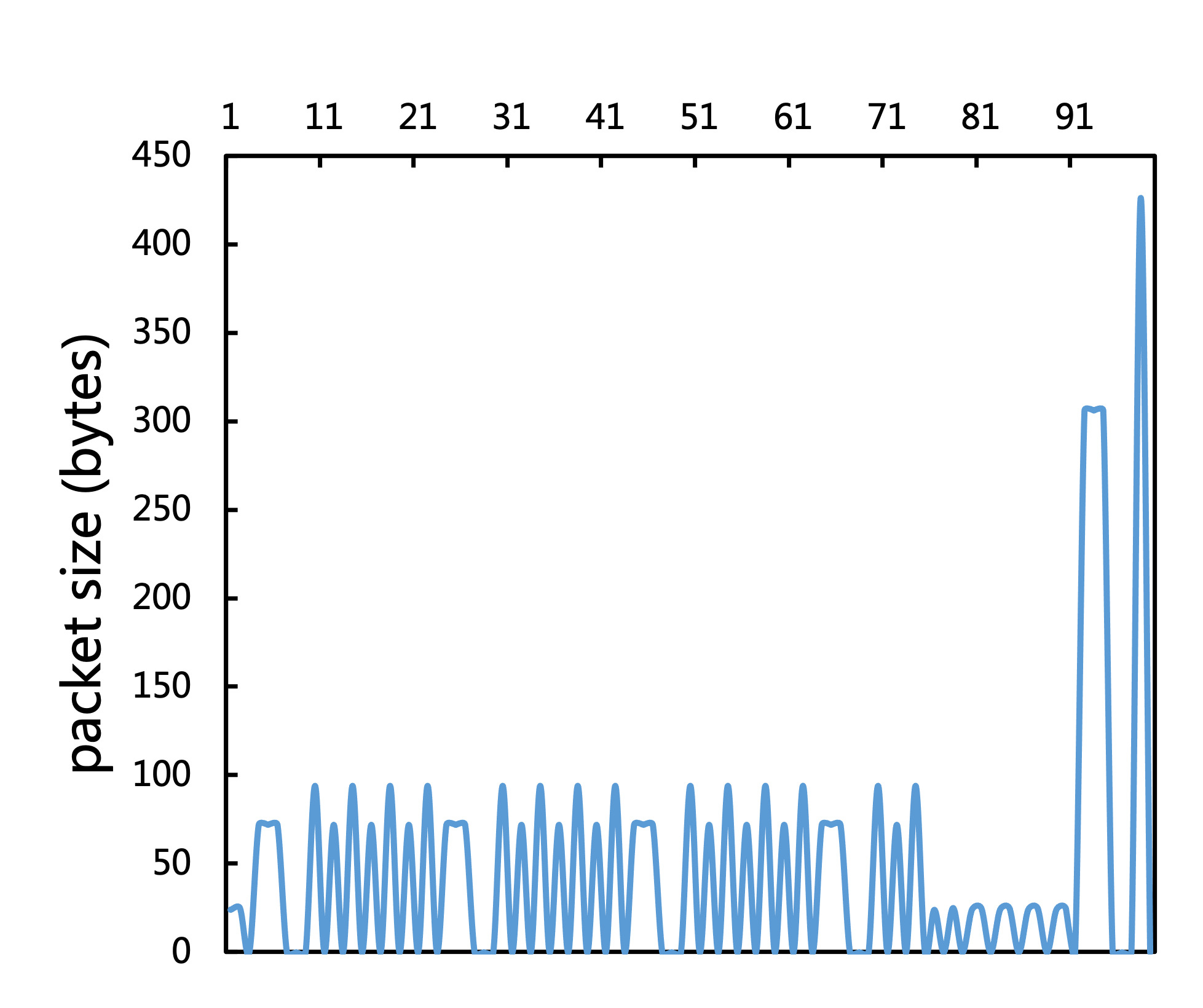}
}\subfigure{d)\label{2 d)}
\includegraphics[width=3.6cm,height=3.4cm]{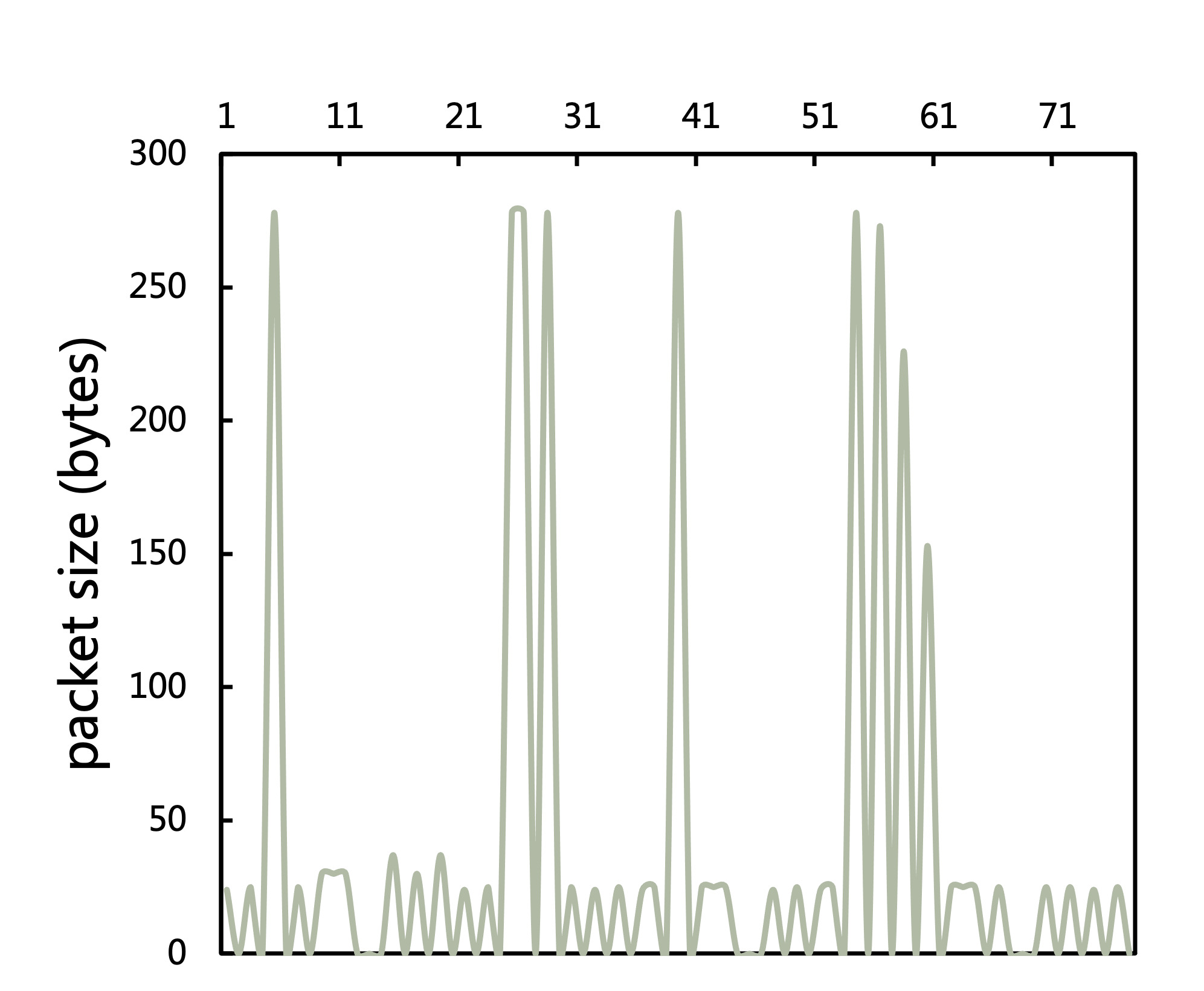}
}
\caption{Comparison of packet sending frequency and packet size of different types of terminals.Fig.\subref{2 a)} shows the comparison of traffic changes of packets sent by three types of terminals within one hour. Fig.\subref{2 b)}, Fig.\subref{2 c)} and Fig.\subref{2 d)} describe the packet size of TTU, LVRC, and LMT terminal samples respectively}    
\label{fig2}       
\end{figure*}

The features are shown in Table \ref{tab1}. The terminals we want to distinguish include the following three types: Low-voltage Meter Reading Concentrator (LVRC), Distribution Transformer Supervisory Terminal Unit (TTU) and Load Management Terminal (LMT). The traffic size of different terminals is distributed as Fig.\ref{fig2}. It can be seen that different terminals have different rhythms in package size and packet sending frequency, so we take the statistical features commonly used in traffic classification methods into consideration. Based on the definition of the flow (each package has the same underlying identity), we have selected several features. 

In addition, we try to mine more identifiable rules of the smart grid data. Since the flow of grid metering terminal is not encrypted, we extracted a byte of information in a specific position of the application layer data that includes 14 states which could represent the behavior of each packet. We count the number of packets that appear in each state according to the two-way transmission and reception, forming a two-way feature of 28-dimensionality. 

\subsection{Hierarchical structure}\label{B}
\begin{figure}[b]
\centerline{\includegraphics[scale=0.65]{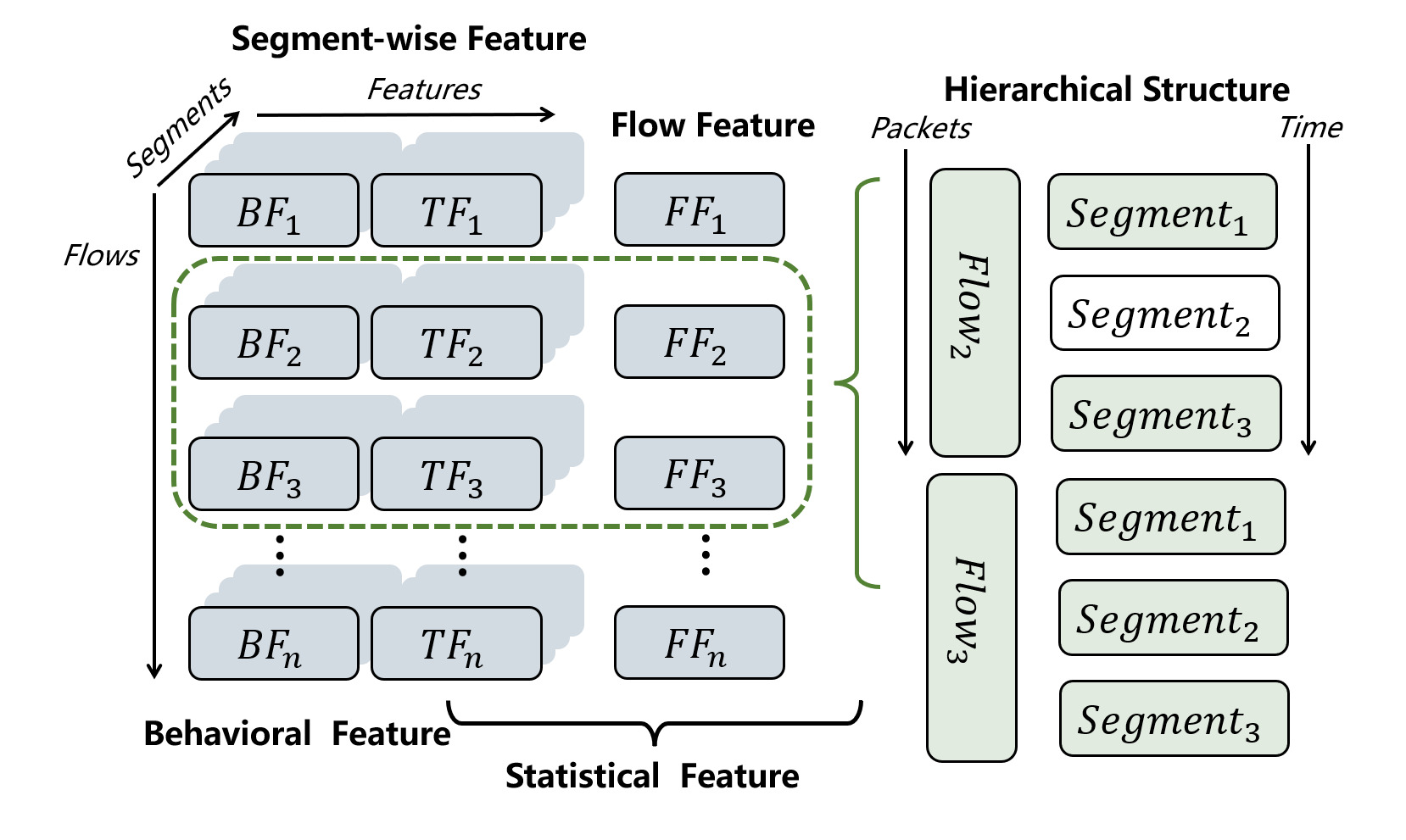}}
\caption{The Hierarchical Structure. \(Flow_2\) and \(Flow_3\) are splitted to 3 segments, but \(Segment_2\) of \(Flow_2\) (marked in white in the figure) is empty as during this time interval no packets were transmitted.}
\label{fig3}
\end{figure}

In order to identify the different states of a network flow over a period of time, we introduce the concept of segment. The proposed hierarchical structure and feature organization strategy are shown in Fig.\ref{fig3}. each flow is divided into \(L\) segments and features are extracted segmentally.  The features of each flow are divided into flow feature and segment-wise feature. Firstly, the statistical features in section \ref{A} are adopted as flow feature. As for the segment-wise feature, we adopt a collection of statistical and behavioral features. The statistical features (TF) in each segment are used to distinguish flows that contain burst traffic, constant sending or receiving data, while behavioral features (BF) are used to discover the behavior sequence of the flow in different time periods. These features can help us discover more fine-grained features.

The segment-wise features (SF) of flow \(F_n\) can be represent as  \(X_{n}^{SF}=\left \{s_{n1}^{SF},s_{n2}^{SF},\cdots ,s_{nL}^{SF}\right \}\in \mathbb{R}^{L\times G}\) where \(L\) is the number of segments and \(G\) is the total feature dimension of each segment. Segment \(l\)'s feature of flow \(F_n\) is \(s_{nl}^{SF}=s_{nl}^{TF}\oplus s_{nl}^{BF}\), where \(s_{nl}^{TF}\in \mathbb{R}^{P}\) represent statistical features and \(s_{nl}^{BF}\in \mathbb{R}^{Q}\) represent behavioral features. Their corresponding feature dimensions are \(P\) and \(Q\), respectively, and \(G=P+Q\). Therefore, our input feature set is composed of  a \(P\)-dimensional flow feature vector and a \(L\times G\)-dimensional segment-wise feature vector, which can be denoted as:
\begin{equation}
\begin{aligned}
\mathbf{X} = \left \{\mathbf{x}^{FF},\mathbf{X}^{SF}\right \},
\end{aligned}
\label{eq0}
\end{equation}
where \(\mathbf{x}^{FF}\) means the flow feature vector set and \(\mathbf{X}^{SF}\) means the segment-wise feature vector set.

\subsection{Model Framework}\label{C}

\begin{figure*}[t]
\centerline{\includegraphics[scale=0.25]{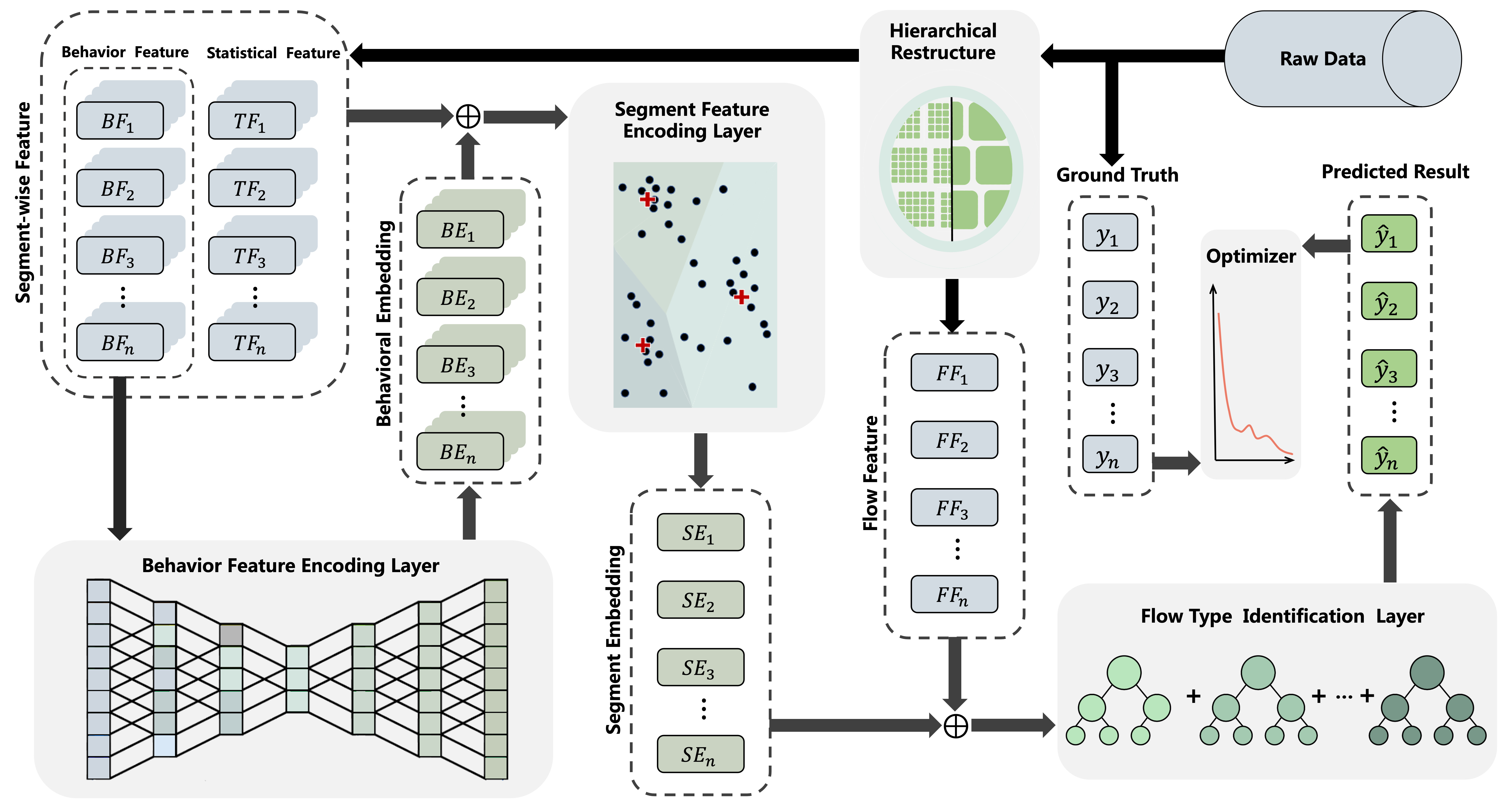}}
\caption{An overall pipeline of the proposed terminal recognize model. The \(\oplus\) indicates that two matrices are concatenated in the feature dimension.}
\label{fig4}
\end{figure*}

As shown in Fig.\ref{fig4}, the proposed model consists of three basic architectures: an autoencoder for encoding behavioral feature set of each segment to obtain potential features, a clustering module for clustering statistical feature set and behavioral embeddings in segments to obtain segment embeddings as potential characterization), and a classification module (utilize the segment embeddings and flow features to recognize terminal type). 

\subsubsection{Behavioral Feature Encoding Layer}

The behavioral feature encoding layer uses an autoencoder to reconstruct the segment-wise behavioral features, replacing the high-dimensional sparse behavioral feature matrix with the compressed representation. In this way, it can be combined with statistical features with reasonable dimensions to jointly describe a segment. We count the number of occurrences of each behavior in the packet according to the two directions of host sending and receiving. Considering some packets whose application layer data is empty, we additionally set a new state ZERO with 2 dimensions for sending and receiving extraction fails. Since most package activities are concentrated in several states, an autoencoder is employed to reconstruct the features, reducing the features while preserving the effective information.

\begin{figure}[t]
\centerline{\includegraphics[scale = 0.3]{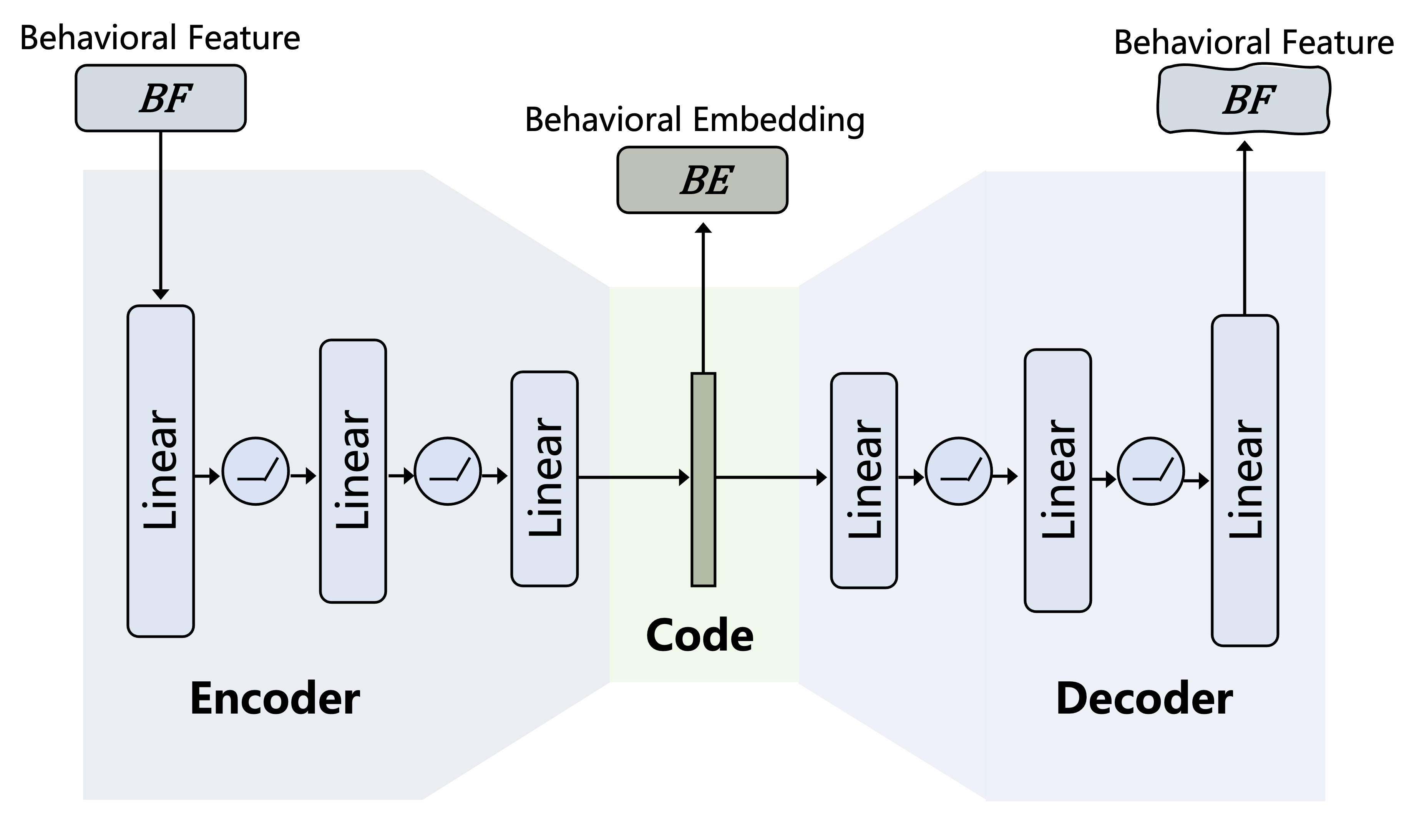}}
\caption{Autoencoder Structure.}
\label{fig5}
\end{figure}

Autoencoder is a type of artificial neural network used to learn efficient coding of unlabeled data. Autoencoder has two main parts: an encoder that maps the input into the encoding, and a decoder that maps the encoding to the reconstruction of the input. The encoding is validated and refined by attempting to regenerate the input from the encoding. The structure of autoencoder is shown in Fig.\ref{fig5}

Consider input segment-wise behavioral features of \(F_n\), \(X_{n}^{BF}=\left \{s_{n1}^{BF},s_{n2}^{BF},\cdots ,s_{nL}^{BF}\right \}\) where \(s_{nl}^{BF}\in \mathbb{R}^{Q}\) is the \(l\)th segment of the flow and \(L\) means the number of segments. The goal of autoencoder is defined to be \(y_{nl}^{BF}=s_{nl}^{BF}\) for \(l\in \left \{1,2,\cdots ,L\right \}\) that \(y_{nl}^{BF}\) is processed by the encoder and decoder as the obtained output. The objective function of an autoencoder can be denoted as \(F_{W, b}\left(S_{n l}^{B F}\right)\) where \(W\) and \(b\) are the whole network weights and biases vectors. The function is used to learn a compressed representation of the input data and approximately learns the identity function:

\begin{equation}
\begin{aligned}
F_{W, b}\left(s_{n l}^{B F}\right) \simeq s_{n l}^{B F},
\end{aligned}
\label{eq1}
\end{equation}
the general form of the loss function of an autoencoder is shown as follows:
\begin{equation}
\begin{aligned}
\mathcal{L}(W, b)=\frac{1}{Q} \sum_{q=1}^{Q}\left|s_{n l q}^{B F}-F_{W, b}\left(s_{n l q}^{B F}\right)\right|^{2},
\end{aligned}
\label{eq2}
\end{equation}
where \(Q\) denotes the dimensions of features in the layer. The output of the encoder is considered as a set of advanced discriminatory features. 

Our autoencoder can be represent as:
\begin{equation}
\begin{aligned}
\begin{array}{l}
s_{n l}^{B E}=\operatorname{En}\left(s_{n l}^{B F} ; \theta_{E n}\right), \vspace{1ex}\\
s_{n l}^{B F^{\prime}}=\operatorname{De}\left(s_{n l}^{B E}, \theta_{D e}\right),
\end{array}
\end{aligned}
\label{eq3}
\end{equation}
where \(s_{n l}^{B F}\in \mathbb{R}^{Q}\)  is the behavioral features of segment \(l\) in flow \(F_n\),  \(\theta_{En}\) and \(\theta _{De}\) represent the parameters of encoder \(\operatorname{En}\left ( \cdot \right ) \) and decoder \(\operatorname{De}\left ( \cdot \right ) \), respectively. The result of behavioral feature encoding layer can be represented as \(s_{n l}^{B E}\in \mathbb{R}^{V}\) where \(V\) denotes the dimension of each segment embedding. In this way, we can obtain behavioral embeddings set \(X_{n}^{B E}=\left\{s_{n 1}^{B E}, s_{n 2}^{B E}, \cdots, s_{n L}^{B E}\right\}\). Empty segments are ignored in this layer.
\subsubsection{Segment Feature Encoding Layer}
In segment feature encoding layer, the data that express the characteristics of each segment consists of two parts, one is segment-wise statistical features, and the other is segment-wise compressed behavioral embeddings.

Our data structure dictates that the data is ingested in flows. Here, we cluster the features in segments. We first describe the input feature set \(\mathbf{H}\) of clustering algorithm:
\begin{equation}
\begin{aligned}
\begin{array}{l}
\mathbf{H}=\left\{H_{1}, H_{2}, \cdots, H_{N}\right\}, \vspace{1ex}\\
H_{n}=X_{n}^{T F} \oplus X_{n}^{B E}\vspace{1ex}\\
=\left\{s_{n 1}^{T F} \oplus s_{n 1}^{B E}, s_{n 2}^{T F} \oplus s_{n 2}^{B E}, \cdots, s_{n L}^{T F} \oplus s_{n L}^{B E}\right\},
\end{array}
\end{aligned}
\label{eq4}
\end{equation}
where \(H_{n}\in \mathbb{R}^{L\times \left ( P+V\right )}\) is the merged result of segment-wise statistical feature set \(X_{n}^{TF}\in \mathbb{R}^{L\times P}\) and behavioral embedding set \(X_{n}^{BE}\in \mathbb{R}^{L\times V}\), \(\bigoplus\) represents the concatenation operation, \(L\) is the number of segments in each flow while \(P\) and \(V\) represent the dimensions of vectors. The input set is \(\mathbf{H}\in \mathbb{R}^{U\times \left ( P+V\right )}\), where \(U = N×L\) denotes the number of all segments.

An unsupervised clustering algorithm is used to cluster the input features of each segment:
\begin{equation}
\begin{aligned}
\begin{array}{c}
\mathbf{z}=\left\{z_{1}, z_{2}, \cdots, z_{N}\right\}={ \operatorname{Cluster} }(\mathbf{H}) , \vspace{1ex}\\
z_{n}=\left\{z_{n 1}, z_{n 2}, \cdots, z_{n L}\right\},
\end{array}
\end{aligned}
\label{eq5}
\end{equation}
where \(\mathbf{z}\in \mathbb{R}^{N}\) is the result of the clustering algorithm. Given a hyperparameter \(K\) for clustering, and cluster the training set by segment in the model training phase to obtain \(K\) clusters. \(\operatorname{Cluster}\left (\cdot\right )=\left \{0,1,2,\cdots ,K-1\right \}\). For the \(l\)th segment of \(F_n\),  \(z_{nl}\) represent the index of cluster after clustering by the algorithm. After clustering, each segment is associated with one of the \(K\) possible clusters indicating the type of flow behavior exhibited in the segment, while a special cluster identification is used to denote the empty segments. 

Segment-wise features are generated by the cluster encoding to which all segments in the flow belong, hereafter referred to as segment embeddings. We perform one-hot encoding on the cluster numbers obtained by clustering each segment in each flow. We then have:
\begin{equation}
\begin{aligned}
\mathbf{x}^{SE}=\operatorname{Onehot}(\mathbf{z})=\left\{x_{1}^{SE}, x_{2}^{SE}, \cdots, x_{N}^{SE}\right\}, 
\end{aligned}
\label{eq6}
\end{equation}
where \(x_{n}^{SE}\in \mathbb{R}^{K+1}\) contains a total of \(K+1\)-dimensional binary codes, each dimension of data indicates whether the cluster exists in the flow, and the extra one dimension is prepared for the empty segment. In this way, we learn segment embeddings as more fine-grained network information.
\subsubsection{Flow Type Identification Layer}
We have the original flow features: \(\mathbf{x}_{n}^{FF}\in \mathbb{R}^{N\times D} \), where \(N\) is the total number of flows and \(D\) is the dimension of flow feature set. The segment embeddings are merged with flow features firstly:
\begin{equation}
\begin{aligned}
\begin{array}{c}
\mathbf{I}=\mathbf{x}^{S E} \oplus \mathbf{x}^{F F}=\left\{I_{1}, I_{2}, \cdots, I_{N}\right\}, \vspace{1ex}\\
I_{n}=x_{n}^{S E} \bigoplus x_{n}^{F F},
\end{array}
\end{aligned}
\label{eq7}
\end{equation}where \(I_{n}\in \mathbb{R}^{K+1+D}\). The feature set of each flow is applied to our supervised learning model. Flow type recognition layer is based on a supervised classification algorithm. A supervised learning model is used and trained with the type of each flow:
\begin{equation}
\begin{aligned}
\widehat{\mathbf{y}}=\left\{\hat{y}_{1}, \hat{y}_{2}, \cdots, \hat{y}_{N}\right\}=\operatorname{Classify}(\mathbf{I}),
\end{aligned}
\label{eq8}
\end{equation}where \(\hat{y}_{n}\) denotes the predicted type of flow \(F_n\). Therefore, the predicted type \(\hat{\mathbf{y}}\) is final result we get, and we use the loss function of the classification algorithm (such as the cross-entropy function) to calculate the difference between \(\hat{\mathbf{y}}\) and the true value \(\mathbf{y}\) to guide the optimization of the classification model.
\section{Experiments and Results}

\subsection{Dataset and basic-features}
For this work, we used real traffic data from the power grid company, which is a good reference for the network traffic classification problem in the field of IoT devices. This dataset contains traces of automated meter terminals installed by the company, collecting terminal traffic in the wireless access area from 17:11-18:11 on January 4, 2021. And the dataset was stored in a .pcap file. The dataset includes 2.3GB of data corresponding to around 35000 flows. We regard the flow of captured syn packets as short links, and found that most end devices maintain long connections for up to an hour. The proportion of long connections reaches 86.1$\%$. Therefore, we choose long connection traffic as our research object. Our data flow comes from the three types of terminals mentioned in Fig.\ref{fig2}, the flow ratio of LVRC, LMT and TTU is 1:1:1.

The pkt2flow utility was used to extract all the flows from the .pcap file. Afterward, the dpkt framework was employed to extract some basic characteristics of each flow. Since the protocol used for the traffic we analyzed was the TCP protocol, so we did not consider UDP network flows. Features are derived from the basic characteristics including timestamp, length of the application layer data, source IP address and port, destination IP address and port, behavior code. Input features at different stages are standardized using the mean and standard deviation of the same features in the current dimension. The standardization operation subtracts the mean and divides the standard deviation for each feature value so that the distribution for that feature presents a mean of 0 and a standard deviation of 1, scaling all features to a similar range, making them comparable with each other.

\subsection{Evaluation metrics }
In multi-classification prediction task we need to identify the types of network flows, where each flow belongs to one of \(N\) classes \(\mathbf{C} =\left \{C_{1},C_{2},\cdots ,C_{N}\right \}\). For predicted results, we separately count the true positives samples \(\operatorname{tp^{n}}\), false positive samples \(\operatorname{fp^{n}}\) and false negative samples \(\operatorname{fn^{n}}\) of each class \(C^n\). Then four evaluation metrics widely used in classification problems are adopted:

\begin{equation}
\begin{aligned}
\begin{array}{c}
\begin{array}{l}
Accuracy= \displaystyle\frac{\sum_{n=1}^{N} \operatorname{tp^{n}}}{\sum_{n=1}^{N}\left(\operatorname{tp^{n}}+\operatorname{fp^{n}}+\operatorname{fn^{n}}\right)},\vspace{1ex}\\
Precision _{macro }= \displaystyle\frac{1}{N} \sum_{n=1}^{N} \frac{\operatorname{tp^{n}}}{\operatorname{tp^{n}}+\operatorname{fp^{n}}},\vspace{1ex}\\
Recall_{macro}= \displaystyle\frac{1}{N} \sum_{n=1}^{N} \frac{\operatorname{tp^{n}}}{\operatorname{tp^{n}}+\operatorname{fn^{n}}},\vspace{1ex}\\
F1-score _{macro }= \displaystyle\frac{2 \cdot Precision _{macro} \cdot Recall_{macro }}{ Precision_{macro }+Recall _{macro }}.
\end{array}
\end{array}
\end{aligned}
\label{eq9}
\end{equation}

\subsection{Models description and parameter setting}
We implement our model as shown in Fig.\ref{fig4}. To evaluate the performance of our proposed model, we firstly compare with models that only use flow features in different classification algorithms, then compare the different unsupervised clustering algorithms in our model. 
\subsubsection{Classification method}
\begin{itemize}
\item LR: Logistic regression classifies samples based on the predictions of a linear regression model, using L1 penalty and liblinear as optimization.
\item RF: Random Forest combines many decision trees into a forest in a random way, and each decision tree determines the final category of the test sample when classifying, the criterion is set to Gini and n-estimators is set to 10.
\item AdaBoost: An ensemble algorithm with decision trees as weak learners. The parameters n-estimators is set to 100 and the learning rate is 1.0.
\item GB: GradientBoost builds an additive model in a forward stage-wise fashion; it allows for the optimization of arbitrary differentiable loss functions. The loss function is set to deviance and the learning rate is 0.1; the number of boosting stages to perform is 200; the adopted criteria is ‘Friedman MSE’ for the mean squared error with an improvement score by Friedman.
\item NN: neural network using two hidden layers with 64 and 16 dimensions.
\end{itemize}
\subsubsection{Clustering method}
\begin{itemize}
\item GMM: A Gaussian mixture model is a probabilistic model that assumes all the data points are generated from a mixture of a finite number of Gaussian distributions with unknown parameters. The number of mixture components is set to 15 and the number of EM iterations to perform is set to 50.
\item K-Means: The K-Means algorithm clusters data by trying to separate samples in n groups of equal variance, minimizing a criterion known as the inertia or within-cluster sum-of-squares. The method for initialization is set to k-means++.
\end{itemize}

At the same time, we change the combination of the segment length t and the number of clusters \(K\) of the cluster coding layer to find the best model. As a result, the segment length \(\tau\) is set to 300s and the number of segments is derived to 12; the cluster \(K\) in segment feature encoding layer is 15. The autoencoder we use employs three hidden layers with 24,16 and 8 neurons in each layer, and we adopt the ReLU as the activation function. Since the autoencoder model is a mirror-symmetric structure, the number of layers, neurons and the activation function in the decoder and the encoder can be mirror images of each other. 

\subsection{Comparison Results}
A total of seven models are compared, five of which are basic models, and the other two are our proposed models for different algorithm combinations. The results of our comparative evaluation experiments are summarized in Table \ref{tab4}. 
\begin{table}[t]
\caption{Experimental Results}
\begin{center}
\resizebox{\linewidth}{!}{
\begin{tabular}{cccccc}
\hline
Model&Algorithm&precision&recall&accuracy&f1-score\\
\hline
Comparison&LR&0.9441&0.9444&0.9443&0.9441\\
model&RF&0.9467&0.9466&0.9468&0.9466\\
&AdaBoost&0.9513&0.9513&0.9512&0.9512\\
&NN&0.9423&0.9425&0.9424&0.9423\\
&GB&0.9764&0.9764&0.9765&0.9764\\
\hline
Proposed&AE+GMM+GB&0.9825&0.9825&0.9827&0.9825\\
model&AE+KMeans+GB&0.9834&0.9833&0.9835&0.9834\\
\hline
\end{tabular}}
\label{tab4}
\end{center}
\end{table}

\begin{figure}[t]
\centerline{\includegraphics[scale=0.25]{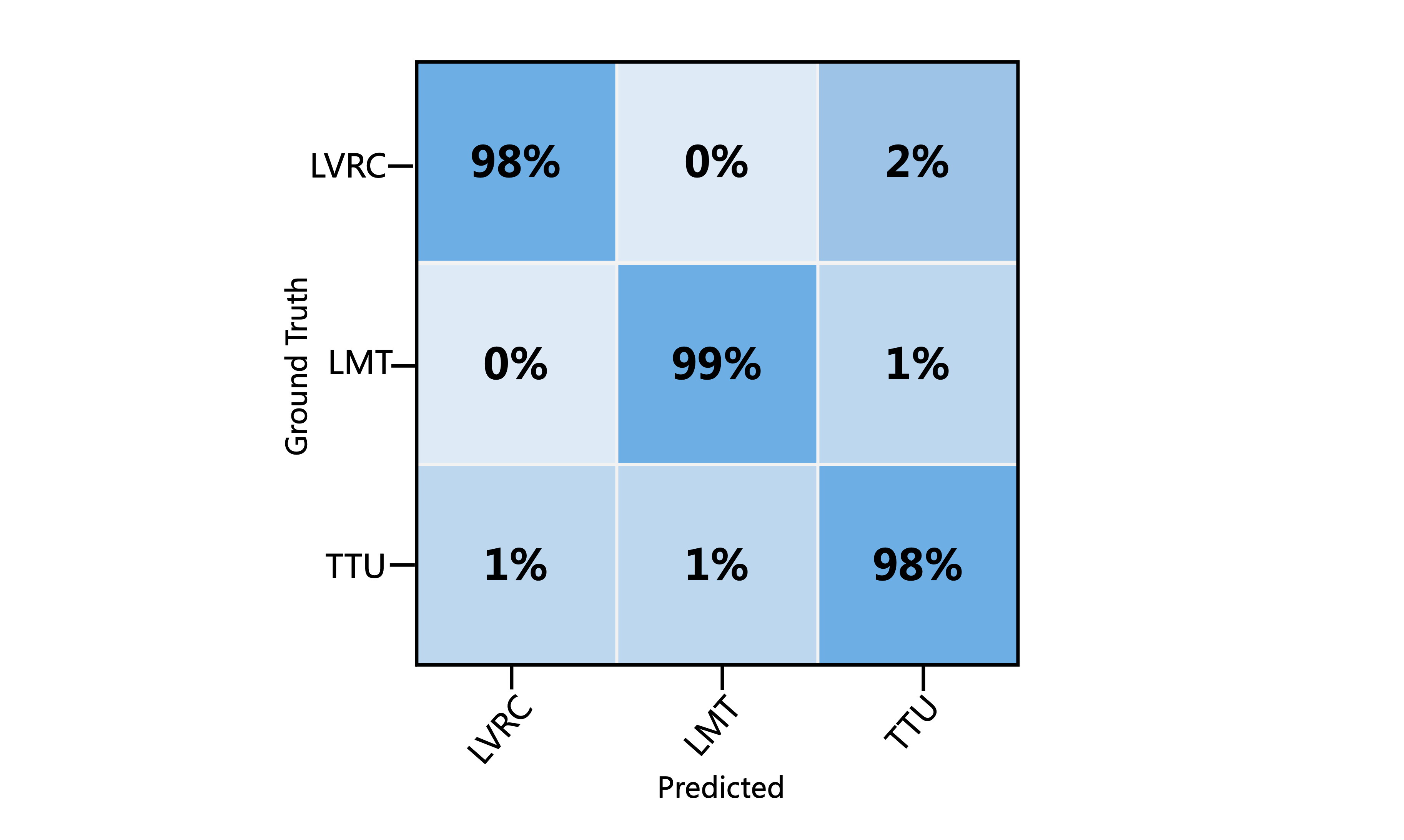}}
\caption{Confusion Matrix of the proposed model.}
\label{fig6}
\end{figure}

Overall, our proposed combined models outperform classification models using only one layer of statistical features, which means the latent features we obtained through behavioral feature encoding layer and segment feature encoding layer are useful for the identification of flows’ type. Among the five comparison models, GradientBoost performs the best, while the other four models perform similarly. When we also employ the GradientBoost classification algorithm, our proposed model outperforms the baseline by 0.7 percentage points in f1-score, and compared with other benchmark models our advantage is even more obvious. As for the proposed model, we also tried different combinations of algorithms. The experimental results show that the autoencoder, K-Means clustering algorithm and GradientBoost classification algorithm achieve the best results, while the f1-score difference is only 0.09 percentage points compared with the GMM clustering algorithm. Therefore, we believe that the main reasons for improving the effect are the following two points: 1) We propose a two-layer structural feature model, so that some hidden features can be discovered through clustering and auto-encoders; 2) We found the most suitable algorithm matching with the feature structure through the comparison of the models. The confusion matrix of the final result obtained by our model is shown in Fig.\ref{fig6}, from which we can observe that our model is very accurate in identifying various grid metering terminals.
\section*{Conclusion}
In this study, we proposed a hierarchical model to recognize the terminal type of traffic flows in the field of smart grids. We construct a set of features including statistical features and behavioral features. In order to make better use of these features, we propose the concept of segmentation, which extracts statistical features and behavioral features in a fine-grained way. Our model employs autoencoders to compress segment-wise behavioral features and uses a clustering algorithm to encode and reconstruct total segment-wise features. Finally, we utilize a classification model to recognize the type of each flow. Through a large number of comparative experiments, we have determined the best algorithm combination. Our final model uses autoencoder, K-Means and GradientBoost, which can effectively recognize the type of grid metering terminals based on network traffic. Extensive experimental analysis shows the performance advantage of our method and confirms that the combination of the feature set, model structure and algorithms contributes significantly to the performance improvement.
\section*{Acknowledgment}

This research is supported by the National Key R$\&$D Program of China under Grant (No. 2019YFB2102400)

\bibliographystyle{IEEEtran}
\bibliography{ref.bib}

\end{document}